\documentclass[twocolumn,pra,aps,floats,eqsecnum]{revtex4}
\usepackage{graphics}
\usepackage[fleqn]{amsmath}
\usepackage{amsxtra}
\usepackage{url}

\newcommand{\etal}{{\it et al.}}

\begin{document}

 \title{Cu NMR Study of Detwinned Single Crystals of Ortho--II  YBCO6.5 }
 \author{Z. Yamani, W.A. MacFarlane\cite{AAA}, B.W. Statt}
\email[e-mail:]{statt@physics.utoronto.ca}
\affiliation{
    Department of Physics, University of Toronto, 60 St. George St.,
    Toronto, ON, Canada M5S 1A7}
 \author{D. Bonn, R. Liang, W.N. Hardy}
\affiliation{Department of Physics and Astronomy, University Of British
Colombia,
 6224 Agricultural Road, Vancouver,
          BC, Canada V6T 1Z1}
 \date{\today}

\begin{abstract}
Copper NMR has been used as a local probe of the oxygen
ordering in Ortho--II
YBa$_2$Cu$_3$O$_{6.5}$  crystals grown in BaZrO$_3$ crucibles.
Line assignments have been made to each of the expected
crystallographically inequivalent sites. The presence of distinct
and narrow lines for these sites as well as the lack of a line
known to be associated with oxygen defects indicates that these
crystals are highly stoichiometric. Our estimate of the lower limit on the
chain length is consistent with that derived from X--ray
diffraction measurements. In addition, we have found no evidence
for static magnetic moments, in contrast to some previous results.
\end{abstract}

 \maketitle

\section{Introduction}

Since the discovery\cite{Muller86} of high temperature
superconductors (HTSC), considerable effort has been devoted to
understanding the electronic and magnetic properties of these
materials at a microscopic level. Although there are a large
number of perovskite HTSC materials, the YBa$_2$Cu$_3$O$_{6+x}$
(YBCO6+$x$) family has a special place. It was the first
superconductor discovered\cite{Wu87} with a transition
temperature $T_c$ above liquid nitrogen temperature with a
relatively simple synthesis route. It is available in almost the
entire region of the phase diagram, easily tuned by varying the
oxygen content in a quasi--continuous way.
Thus it is appealing for
probing the intrinsic properties of HTSCs. In the underdoped
regime, a variety of crossover phenomena are observed at
temperatures above $T_c$ in which various forms of spectral weight
at low energies are suppressed. These phenomena are associated
with the opening of a pseudogap\cite{statt99}. To help
elucidate the nature of this pseudogap, it is desirable to have highly
stoichiometric crystals with a high degree of crystalline
order. In particular, such samples are needed if one wants
to determine whether or not the crossover behaviors are
transistions smeared by disorder and/or impurities\cite{chakraverty01}.

YBCO6+$x$ can be prepared in the whole range of oxygen content $6+x$
= 6.0 to slightly larger than 7.0. The oxygen content affects the
crystallographic structure of the material\cite{Ginsberg89}. At
$x$=1.0, the crystal structure consists of a bilayer of
two--dimensional (2D) CuO$_2$ planes and a single layer of
one--dimensional (1D) Cu--O--Cu chains, hereafter denoted the CuO
chain layer. Nuclear magnetic resonance (NMR) spectroscopy offers
a unique probe to investigate the microscopic properties of the
CuO$_2$ planes and CuO chain layers separately. In particular, the
response of inequivalent sites can in principle be resolved in NMR
spectra and the dynamical properties of these sites studied by
nuclear relaxation. The chains are important not only because of
their role as charge reservoirs and their proximity to the
superconducting CuO$_2$ planes, but also because of their low
dimensionality. It is well known that charge/spin density
wave(CDW/SDW) ground states are prevalent in quasi--one
dimensional systems\cite{grunerbook}. Therefore, the chains in
this system present an important opportunity for clarifying the
question of whether  such transitions exist in HTSC. In fact this
has been the subject of study in the stoichiometric compound
YBa$_2$Cu$_4$O$_8$ containing two CuO chain layers, where evidence
for the presence of incommensurate CDW fluctuations, near the wave
vector $2k_F$ occurring below 180 K has been
reported\cite{suter97}. There are a few reports on charge
modulations in
 YBCO7 as well\cite{grevin00}. 
Perhaps the most striking evidence for 
charge oscillations in the chains of YBCO7 is provided
by STM observations\cite{derro02}.

The structural phase diagram of YBCO6+$x$ has been studied in great
detail; early neutron diffraction reveal\cite{jorgensen87} a
tetragonal to orthorhombic phase transition as the sample is
cooled from the growth temperature, in which oxygens form 1D CuO
chains and thus break the tetragonal symmetry. Typically the chain
ordering occurs in domains and the samples need to be mechanically
detwinned in order to produce crystals with a single orientation
of chains.

\begin{figure}[!]
\begin{center}
\resizebox{1.0\linewidth}{!}{\includegraphics{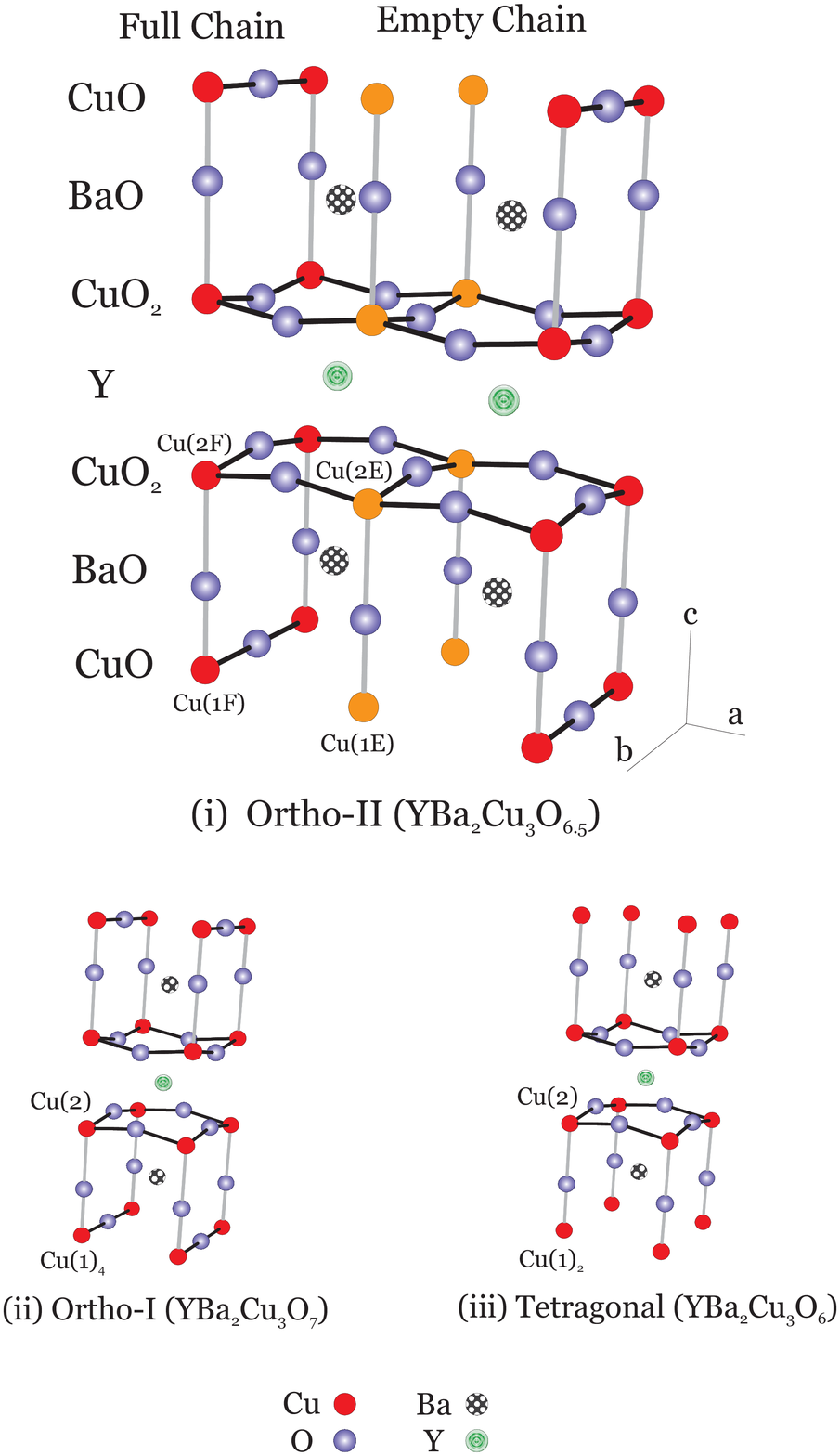}} \vskip
0.1cm
\end{center}
\caption[]{
  The crystal structure of YBCO6+$x$ for (i) x=0.5 (Ortho--II) (ii) x=1.0 (Ortho-I)
  and (iii) x=0.0 (T).} \label{structure}
\end{figure}

Depending on the preparation conditions and oxygen content, oxygen
atoms arrange themselves in different types of ordering in the CuO
chain layers. Efforts to produce crystals with domains
of long chains requires an understanding of oxygen ordering in the 
chain layer. In addition to the orthorhombic structure (Ortho--I,
full chains) at $x$=1.0 (Fig. \ref{structure} (ii)) and tetragonal
phase (T, empty chains) at $x$=0.0 (Fig. \ref{structure} (iii)),
at least four more modified orthorhombic structures have been
observed\cite{Zimmeretal99} such as Ortho--II (Fig.
\ref{structure} (i)) for 6.35 $\le$ 6+$x$ $\le$ 6.62 with
alternating full and empty chains. There have been theoretical
studies based on lattice gas models to explain the structural
phase diagram of YBCO6+$x$\cite{asynnni}. These models can be shown
to be equivalent to the Ising model with asymmetric next nearest
neighbor interactions (ASYNNNI model) which use three pair
potentials between oxygen atoms in CuO chain layers: $V_1$ for
nearest oxygen neighbor pairs, $V_2$ for next nearest pairs
connected through a Cu ion and $V_3$ for those without Cu in
between, with the condition $V_2<0<V_3<V_1$. It is the strongly
repulsive interaction $V_1$ of two \textit{inter-chain} nearest
neighbor oxygen ions which prevents the occupation of nearest
neighbor oxygen sites and promotes the formation of chains. Long
chain segments are induced by the attractive potential $V_2$
between two \textit{intra-chain} next nearest neighbor oxygen ions
connected through a Cu ion. Alternating sequences of full and
empty chains follow from the weak repulsive $V_3$ between two
\textit{inter-chain} next nearest neighbor oxygen ions.

In YBCO6+$x$, $T_c$ changes from the maximum 92.5 K at optimal
doping to zero for fully oxygen depleted samples\cite{kaldis01}.
This change takes place in a nonlinear manner with a  pronounced
plateau at  60 K for 6.45 $\le$
 6+$x$ $\le$ 6.65 and a
broad maximum at 92.5 K for 6.85 $\le$ 6+$x$ $\le$ 7.0. These
features are associated with the Ortho--II and Ortho--I ordered
phases, respectively. It is suggested~\cite{cava87} that this
correlation has to do with charge transfer rather than with any
specific $T_c$ of the ordered phases. The importance of oxygen
ordering in the CuO chain layers in determining the charge
transfer was first noted in a study of the time dependence of
$T_c$ during room temperature annealing of quenched
samples\cite{Kwok88}. With the oxygen content (average occupancy
of different oxygen sites) remaining constant, the degree of
oxygen ordering increased with time. This resulted in
an increase in the orthorhombicity and in $T_c$. It was shown that
the rearrangement of oxygen ions in the chain layers to form
longer chain segments was responsible for this increase. Another
experimental indication of the presence of a minimum chain length
for the charge transfer to take place is the $T_N$ vs. oxygen
content for small $x$: $T_N$ is nearly constant for
6+$x\le$6.2~\cite{TNx}. The relationship between oxygen
ordering and hole concentration has been addressed\cite{zaanen88}
theoretically by band structure calculations. Chain fragments were
found to be unable to transfer charge to the planes until they
exceeded a critical length equal to three (containing three Cu and
two O atoms). An alternative model combining the ASYNNNI and the
charge transfer models has also been\cite{poulsen91} successful in
explaining the variation of $T_c$ vs. $x$. Here it is assumed that
only the ordered phase participates in superconductivity. Thus
there is a minimum domain size below which no charge transfer can
take place.

NMR experiments can be used to measure the electric field gradient
(EFG) in materials. Since the EFG at a nucleus
depends on the symmetry of that site, the valence of the ion
itself and the charge of its environment, valuable information
about the electronic structure of the system under study can be
obtained by such measurements; i.e. charge transfer. However, the interpretation
can be complicated if the samples are not single phase
or suffer from impurities and/or disorder. There are a number of
NMR 
reports on underdoped YBCO6+$x$ with different oxygen
contents\cite{Horvatic93,Yasuoka89,Tornau94,Lutgemeier92,Warren89}
where either the sample is not stoichiometric or suffers from
oxygen inhomogeneity, powder averaging or twinning.
L\"{u}tgemeier's group\cite{Lutgemeier92} was the first to study
the effects of oxygen ordering on the NMR properties of YBCO6+$x$.
However, they used aligned powdered samples where the NMR
resonances were broadened by non--stoichiometry and oxygen disorder. 
For Ortho--II ordering only a
partially resolved splitting of the resonance line from the
CuO$_2$ plane was observed.

With recent enhancements in crystal growth very pure and highly
ordered detwinned Ortho--II crystals with large oxygen--ordered domain
size have become available\cite{Bonn00}. To probe the intrinsic
properties of the Ortho--II phase of YBCO we have performed NMR
experiments on highly ordered detwinned Ortho--II single crystals
of YBa$_2$Cu$_3$O$_{6.5}$ (YBCO6.5). We report NMR spectra and
line assignments for these Ortho--II crystals together with NMR
evidence pertaining to the quality of the crystals, specifically
regarding the chain lengths and the lack of oxygen
disorder. We also report EFG parameters for different Cu sites in
Ortho--II structure along with an analysis regarding lattice and
valence charge contributions. We will discuss the relation between
oxygen ordering and charge transfer in the Ortho--II structure. It
will be shown that these crystals are an extremely important
disorder--free prototype of the underdoped state of HTSC.

\section{Experiment}\label{exp}

Single crystals of YBCO6.5 were grown by a flux method using
BaZrO$_3$ crucibles\cite{Bonn00}. The crystals were first
detwinned and then the oxygen content set to 6.50 during the final
annealing process. The NMR properties of two such crystals used in
our experiments proved to be identical to within experimental
error. Hence we will not distinguish between them. The crystals
were 2 mm $\times$ 2.5 mm wide and $\sim$ 0.1 mm thick platelets with the
$c$--axis perpendicular to the platelet and $a$ and $b$ axes
parallel to the edges of the platelet. The superconducting
transitions of the crystals measured by a SQUID magnetometer give
$T_c$ = 62 K with a transition width $\Delta T_c$ = 0.6 K. X--ray
diffraction measurements performed on these crystals show a
three--dimensional Ortho--II oxygen ordering with long correlation
lengths 148, 430 and 58 \AA\ along the $a$, $b$ and $c$
crystallographic axes, respectively\cite{Bonn00}. The long
correlation length along the $b$ (chain) axis indicates the
presence of much longer chains than have been available
previously\cite{Lutgemeier92}. Both the large oxygen--ordered
domain size and a small $\Delta T_c$ are indicative of high degree
of oxygen homogeneity in these crystals. Because twin boundaries
promote oxygen clustering, the removal of twin boundaries was
found to be essential for obtaining a highly ordered Ortho--II
phase\cite{Bonn00}. Single crystal X--ray diffraction measurements
on our crystals showed at most 10\% of the sample was
twinned.

Cu NMR data were taken between 30 and 250 K and magnetic fields up
to 9 T with a home built pulsed spectrometer with quadrature
detection. The crystal was placed in either a Pt or Au rectangular
coil tunable from 30 to 100 MHz. A rectangular coil was chosen to
conform to the shape of the crystal and so
maximize the filling factor. The onset of the superconducting
transition of the sample was observed \textit{in situ} at 62 K by
a sharp change in the inductance of the tuned coil at zero applied
magnetic field. A powder Al reference sample was located in the
coil adjacent to the sample to calibrate the value of the magnetic
field at the sample position.

The data were obtained with a standard two--pulse spin--echo
sequence, with 4 $\mu$s pulse length and 30 to 40 $\mu$s pulse
separation. Copper nuclei have two isotopes $^{63}$Cu and
$^{65}$Cu with spin $\frac{3}{2}$. Each isotope gives one central
transition ($\frac{1}{2}$ $\leftrightarrow$ $-\frac{1}{2}$) and
two satellites ($\pm\frac{3}{2}$ $\leftrightarrow$
$\pm\frac{1}{2}$). To obtain the complete Cu NMR spectrum of the
sample we used either field or frequency sweep methods. The
spectra were then constructed by superimposing the Fourier
transform spectra of the spin echo measured at a certain frequency
or field interval. There was no attempt to correct the signal
intensity for $T_2$ or frequency dependencies in frequency sweep
measurements, given the latter's narrow range. Although the
samples are high quality, because of their small volume a
significant amount of signal averaging was required to obtain a
satisfactory signal to noise (S/N) ratio.

\section{Results and Discussion} \label{R&D}

\subsection{Line Assignment}\label{LineAss}

\begin{figure}[tbh]
\resizebox{1.\linewidth}{!}{\includegraphics{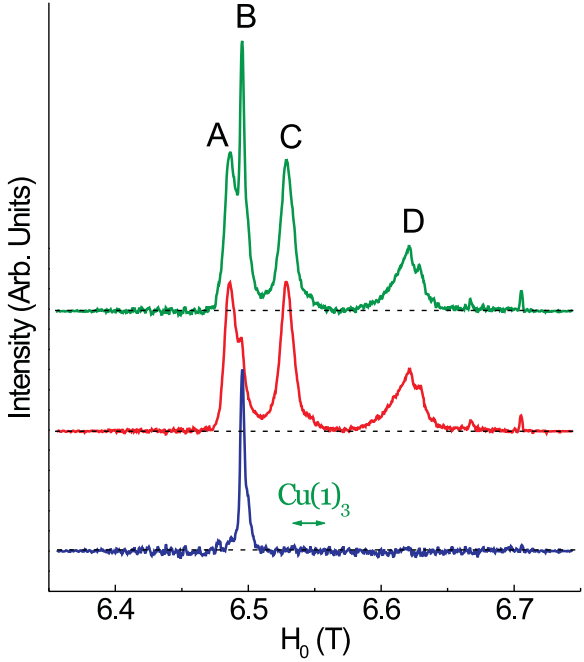}} 
\vskip -0.5cm \caption[]{$^{63}$Cu NMR spectra for central
transitions at 60 K and 75.75 MHz for {\bf H}$_0||${\bf a} taken
at different repetition time, T$_{rep}$ (top panel: T$_{rep}$=250
msec, middle panel: T$_{rep}$=25 msec, and lower panel: the
difference). The horizontal arrow indicates the possible range
for Cu(1)$_3$. See
text for details. The two narrow and small signals in the far
right side of the spectra are from Cu metal in the sample probe.}
\label{spec1}
\end{figure}

Typical $^{63}$Cu NMR spectra of central transitions obtained
at 60 K at 75.75 MHz for the applied field parallel to the $a$--axis
({\bf H}$_0||${\bf a}) are shown in Fig.~\ref{spec1}. Similar
spectra were also observed for {\bf H}$_0||${\bf b} containing
four resonance lines but at different positions. We denote the
observed lines in Fig.~\ref{spec1} by A, B, C and D. 
The spectrum for A--D resonances has also been
obtained for $^{65}$Cu with intensities consistent 
with the natural abundances of the two isotopes.
These four Cu
resonances are associated with the four inequivalent Cu sites in the
unit cell as expected from the Ortho--II crystal structure
(Fig.~\ref{structure}(i)). 
Two lines A
and C appear nearly with the same width, integrated area and
relaxation rates. Line B has a very long spin--lattice relaxation
time (nearly three order of magnitude longer than A and C) 
and is relatively narrow compared to the rest of the lines.
Line D is broader compared to line A and C and has an order of
magnitude longer
relaxation time. It should be noted that these resonance lines
are well resolved and that there appears to be no background
signal under these lines. Also note that a relatively long repetition
time of 250 msec must be used in order not to saturate
line B. Most of the spectra are collected with a shorter
repetition time of 25 msec in order to saturate line B which
otherwise obscures line A.

Line assignments will be made by comparing the local parameters
of the four Cu sites with those of YBCO6 and YBCO7. To motivate
this comparison consider the structural environmemt of each Cu site.
Ortho--II YBCO consists of alternating full and empty chains (see
Fig.~\ref{structure}(i)) and hence there are two inequivalent Cu
sites in CuO chain layers, one in full Cu(1F) and one in empty
Cu(1E) chains. Correspondingly there are two inequivalent Cu sites
in the CuO$_2$ planes, adjacent to full and empty chains Cu(2F)
and Cu(2E), respectively. Comparing the crystal structure of
Ortho--II with that of YBCO7 (Fig.~\ref{structure}(ii)), one
concludes that the local structural environments of Cu(2F) and
Cu(1F) are similar to that of Cu(2) in the plane and four--fold
coordinated Cu in the chain layers Cu(1)$_4$ in YBCO7,
respectively. Similarly, the Cu(2E) and Cu(1E) sites have the same
local structural environment as  Cu(2) in the plane and two--fold
coordinated Cu in the chain layers Cu(1)$_2$ in YBCO6.  NMR
parameters for these lines will be compared  with the
known\cite{penn88,shimizu88,mali91} NMR parameters (Knight shifts,
nuclear quadrupole frequency and asymmetry parameters) of the
corresponding sites in YBCO7 and YBCO6 to determine the line
assignments.

\begin{figure}[tbh]
\resizebox{1.\linewidth}{!}{\includegraphics{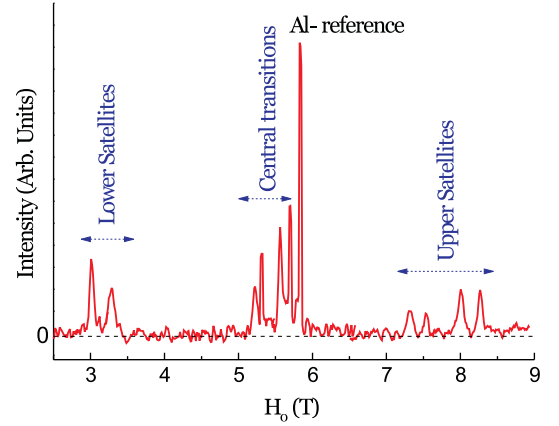}} 
\vskip -0.5cm \caption[]{ The complete Cu NMR spectrum at 70 K and
64.85 MHz for {\bf H}$_0||${\bf c}. For this direction, the
central transitions and lower satellites of A and C lines
overlap.} \label{totspeccax}
\end{figure}

In order to obtain the NMR parameters for the observed lines, we
have measured the complete $^{63/65}$Cu spectrum including the
satellite and central transitions at different frequencies and
fields along the three crystallographic axes. A typical complete
spectrum is shown in Fig.~\ref{totspeccax} for {\bf H}$_0||${\bf
c} at 70 K. Note that for this direction of the applied field, the
central and lower satellite transitions of lines A and C overlap.
The spectra were analyzed using second order perturbation theory
for the quadrupolar shift to the Zeeman interaction (see Appendix)
for central transitions and by exact diagonalization for {\bf
H}$_0$ parallel to crystallographic axes for the satellites since
second order perturbation theory is insufficient for these
resonances.

\begin{figure}[tbh]
\resizebox{1.\linewidth}{!}{\includegraphics{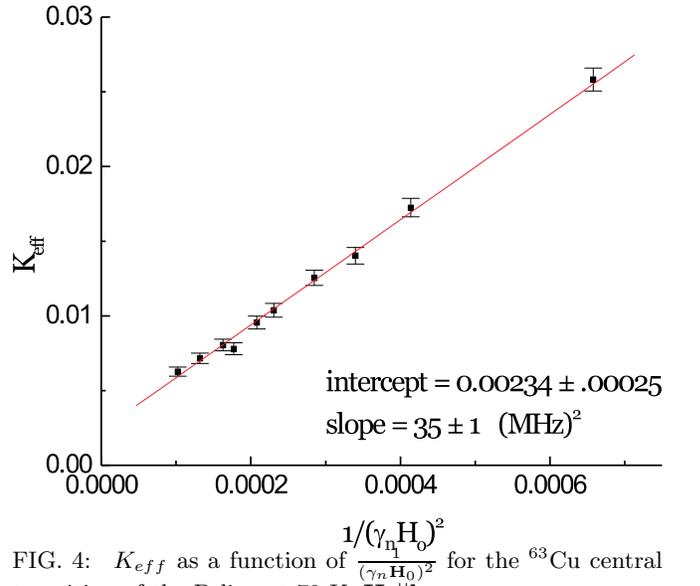}} 
\vskip -0.5cm \caption[]{ $K_{eff}$ as a function of
$\frac{1}{(\gamma_n {\bf H}_0)^2}$ for the $^{63}$Cu central
transition of the D line at 70 K, {\bf H}$_0||${\bf b}.}
\label{keff}
\end{figure}

If one writes the resonance frequency $\nu$ (see Appendix) of the
central transition to the second order as
\begin{equation}\label{SOP}
\frac{\nu}{\gamma_n H_0}-1 =K_{eff}= K +
\frac{f(\nu_Q,\eta,\theta,\phi)}{1+K} \frac{1}{(\gamma_n H_0)^2}
\end{equation}
it is easily seen that $K_{eff}$ is a linear function of
$\frac{1}{(\gamma_n H_0)^2}$ with the intercept equal to $K$ and the
slope proportional to the quadrupolar coupling constant for that
direction.  For example, in Fig.~\ref{keff}, we have shown this
data for the central transition of the line D obtained at 70 K and
for {\bf H}$_0||${\bf b}. We have used the position of the center
of gravity of this line to obtain $K_{eff}$ since this line is
broad at low temperatures.  The Knight shifts and EFG
parameters are obtained for all sites with a simultaneous fit
of the central transition and satellite data in the three field 
orientations\cite{crc}.

The NMR parameters for the four observed lines in the normal
state are reported in Table~~\ref{t:nmrpara}. For comparison we
have listed the NMR parameters of Cu(2) and Cu(1)$_4$ sites in
YBCO7 and Cu(2) and Cu(1)$_2$ sites in YBCO6 in
Table~\ref{t:summary}\cite{penn88,shimizu88,mali91}. We can
readily assign the four resonances observed in Ortho--II to the
four different crystallographic sites by comparing their NMR
parameters to those of Cu sites in YBCO7 and YBCO6. This
comparison shows that the two lines A and C, with very similar
features, come from the planes; A from the planar Cu sites
adjacent to the full chains Cu(2F) and C from the planar Cu sites
adjacent to the empty chains Cu(2E). The EFG principal axis for
both sites is along the $c$--axis which is the same for the Cu(2)
sites in YBCO7 and YBCO6. The asymmetry parameter for these sites
is very small (nearly axially symmetric) as with the Cu(2) sites
in YBCO7 and YBCO6. The slightly larger asymmetry parameter
observed in Ortho--II might be explained by crystal structure
arguments since Ortho--II structure has a lower symmetry than the
Ortho--I or tetragonal phases. The Cu(2F) and Cu(2E) resonances
appear at different positions because of their different quadrupolar coupling
constants. Since the quadrupolar coupling depends on the ion's
surrounding charge distribution as well as the local charge at
the position of the ion, a different coupling is expected for the
two planar Cu sites as observed. We will discuss the contribution
of the lattice and valence electrons to this coupling in more
detail below. The $\nu_Q$ for Cu(2E) is higher than the observed
one in YBCO6, consistent with the trend of $\nu_Q$ vs. 
doping\cite{kumagai01}. The same intensity of the Cu(2F) and Cu(2E)
lines\cite{T2comment} within our S/N is an indication 
that there are equal numbers
of these sites in the crystal. Hence there is a high degree of
ordering in the chain layers with a periodic sequence of full and
empty chains. The mere fact that we have observed only two
distinct Cu(2) signals for both satellite and central transitions
(Fig.~\ref{totspeccax}) is suggestive of a complete ordering of
the full and empty chains. In a disordered structure the signal
would merge to one broad resonance with some splitting depending
on the degree of ordering. Such partially resolved splittings for
the planar Cu resonances is typical\cite{Lutgemeier92} in previous
generations of YBCO6+$x$ samples.

\begin{table*}
\caption{Experimental NMR parameters for the four observed lines
at 70 K. Knight shifts are reported in units of percent and
$\nu_Q$ in MHz. EFG--PA represents the direction of $V_{zz}$. Site
assignment is based on the comparison of these parameters with
those of Cu sites in YBCO7 and YBCO6. For details see the text. }
\begin{ruledtabular}
\begin{tabular}{llcccccc}\label{t:nmrpara}
  resonance &site & K$_a$ & K$_b$ & K$_c$ & $\nu_Q$ & $\eta$ &  EFG--PA\\
\hline
  A &Cu(2F)& 0.25$\pm$0.01 & 0.25$\pm$0.01& 1.18$\pm$0.01& 30.55$\pm$0.03    & 0.018$\pm$0.002     &c\\
  B &Cu(1E)& -0.4$\pm$0.1& 0.07$\pm$0.05& -0.1$\pm$ 0.1& 32.1 $\pm$0.2    & 0.00$\pm$0.01     &c\\
  C &Cu(2E)& 0.26$\pm$0.01& 0.24$\pm$0.01& 1.14$\pm$0.01& 27.43$\pm$0.03    & 0.022$\pm$0.002     &c\\
  D &Cu(1F)& 0.9$\pm$0.1& 0.23$\pm$0.03& 0.23$\pm$0.05& 19.6 $\pm$0.1    & 0.8$\pm$0.1     &a\\
\end{tabular}
\end{ruledtabular}
\end{table*}

\begin{table*}
\caption{NMR parameters for Cu(2) and Cu(1)$_4$ sites in YBCO7 and
Cu(2) and Cu(1)$_2$ sites in YBCO6 materials. Knight shifts are in
units of percent, $\nu_Q$ in MHz. EFG--PA represents the direction
of $V_{zz}$.}
\begin{ruledtabular}
\begin{tabular}{llcccccc}\label{t:summary}
& site &K$_a$& K$_b$& K$_c$& $\nu_Q$ & $\eta$& EFG--PA\\
\hline YBCO7\footnotemark[1]   & Cu(2)     & 0.59$\pm$0.04 &
0.59$\pm$0.04 &
1.267$\pm$0.001& 31.54$\pm$0.05 &0.01$\pm$0.01 &c \\
                           & Cu(1)$_4$ &  1.38$\pm$0.07 &
0.55$\pm$0.07 & 0.60$\pm$0.04& 19.35$\pm$0.08 &0.95 $\pm$0.02 &a\\
\hline

  YBCO6\footnotemark[2]   & Cu(2) & 0.47$\pm$0.02 &
  &1.335$\pm$0.005
& 23.80 $\pm$0.2&0&c\\
          & Cu(1)$_2$  & 0.18 $\pm$0.01&  &-0.117$\pm$0.003 &
          29.54
$\pm$0.02&0&c \\
\end{tabular}
\end{ruledtabular} \footnotemark[1]{The Knight shifts are from
Ref.\cite{penn88}, $\nu_Q$ and $\eta$ are taken from Ref.
\cite{shimizu88} all at 100 K. } \footnotemark[2]{Taken from
Ref.\cite{mali91} at 505 K.}
\end{table*}

The resonance line B appears in the same frequency range as Cu(2F)
but can be distinguished by its much longer relaxation time. In
fact, among the four observed resonance lines, B has the smallest
width, Knight shift and the longest relaxation time (of order
seconds) comparable with that of Cu(1)$_2$ in the insulating
tetragonal phase of YBCO6\cite{pozzi99}. A comparison of the NMR
parameters of this line with Cu(1)$_2$ present in YBCO6 with empty
chains leads us to assign this line to Cu(1E) in the Ortho--II
structure. Similar to Cu(1)$_2$, the Cu(1E) resonance has axial
symmetry. The long spin--lattice relaxation time is interpreted as
a result of a low conduction electron density at the resonant
nucleus. This is consistent with the Cu ions being in the
3d$^{10}$ electronic configuration. A small Knight shift is also
consistent with a low density of conduction electrons at this
site. In addition, the fact that the resonance signal from this
site is very narrow  is in agreement with this conclusion because
insulating sites tend to be much more uniform in nature than
conducting ones as they lack an inhomogeneous $K_s$ component. The
Cu(1E) linewidth is 40 kHz at 60 K with {\bf H}$_0||${\bf a} = 6.5
T for different directions of applied field and becomes even
narrower at higher temperatures.

Line D is broader than other resonances, at least at low
temperatures. We have found a large value for its asymmetry
parameter (see Table~\ref{t:nmrpara}). The highly asymmetric
behavior of the EFG is reminiscent of Cu(1)$_4$ in YBCO7. This
fact together with its similar $\nu_Q$ has led us to assign this
line to the Cu(1F) site. An examination of Cu(1F) site symmetry
explains its highly asymmetric nature. In the Ortho--II structure,
the Cu(1F) sites are surrounded by four oxygen ions, two chain
oxygens and two apical oxygens. Since these oxygen neighbors are
not all equivalent, a non-symmetric EFG is expected. To understand
the nature of the broadening of this line at low temperatures, we
have studied the line shape and relaxation rates for several
directions of the applied field. We attribute this broadening to
the presence of Friedel--like oscillations nucleated at the chain
ends. The details of this study will be presented
elsewhere\cite{zyamaniChain}.

We have confirmed that the ratio of quadrupolar shifts of the two isotopes
measured at the same site differs by their quadrupole moment
($^{63}$Q/$^{65}$Q = 1.079\cite{crc}) for all sites.
This indicates that no internal magnetic field, due to static
magnetic moments, is present at these Cu sites since such a field
would shift or broaden the lines. This is in disagreement with
some neutron--scattering measurements where a moment of $\sim$
0.05 $\mu_B$ is observed\cite{sidis} in YBCO6.5 and of $\sim$ 0.02
$\mu_B$ in YBCO6.6\cite{mook} for copper sites in the CuO$_2$
planes, and with $\mu$SR results on YBCO6.67 showing\cite{sonier1}
the existence of static magnetism in these planes. A magnetic
moment of the order of $\sim$ 0.02 $\mu_B$ would have shifted or
broadened the Cu(2F/E) lines by about 3 MHz which we would have
easily observed. More recent neutron scattering experiments on
Ortho--II found\cite{stock} no evidence for the presence of static
magnetic moments. Also, in a further work, it was
shown\cite{sonier2} that the $\mu$SR results might arise from
charge inhomogeneity rather than magnetic order.

To further elucidate this apparent discrepency several factors should
be noted. As reported by Sidis \etal\cite{sidis} their static moments
fluctuate on a nanosecond time scale which would average out these
moments on the NMR time scale. However this still does not settle
the discrepancy with the work of Stock \etal\cite{stock}. In this
experiment the Ortho--II sample is much more stoichiometric than
that of Sidis \etal or the YBCO6.6 sample of Mook \etal. Thus, as
suggested by Stock \etal, oxygen disorder in the chains may be
responsible for establishing magnetic order there. With respect
to the $\mu$SR results, Sonier \etal\cite{sonier2} have suggested
that for samples with $x >$ 6.5 local charge inhomogeneity,
caused by CDW in the chains as shown by  Gr\'{e}vin \etal\cite{grevin00}
could lead to magnetic moments via the stripe mechanism.
However for the Ortho--II concentration of $x$ = 6.5 no $\mu$SR
signature of a CDW was detected. From our work
on the Cu(1F) line mentioned above\cite{zyamaniChain} we 
have established that a Friedel--like spin--density oscillation exists in
the chains of sufficient magnitude to produce the field, of
order 1 gauss, necessary to account for the Ortho--II $\mu$SR results. Thus
the CDW/stripe mechanism need not be invoked in the case of
stoichiometric Ortho--II YBCO.

\subsection{Chain Length}

We now estimate the chain lengths from our NMR data. In YBCO6+$x$
compounds with intermediate oxygen content, a resonance line with
$\nu_{NQR}\simeq$ 23 MHz is observed\cite{Lutgemeier92} and is
associated with three--fold coordinated Cu(1)$_3$ sites
corresponding to the terminal Cu ion of the full chains with a
relative abundance that diminishes with increasing length of the
chain segments. A similar value of $\nu_Q$ for Cu(1)$_3$ and
Cu(1F) sites together with the fact that Cu(1F) is very broad at
low temperatures, makes it difficult to obtain a precise value for
the concentration of these sites in our sample. As discussed above
we expect the concentration of Cu(1)$_3$ to be low. In fact, in the
complete NMR spectrum for Ortho--II measured between 1 to 9 T, to
within our S/N we did not observe any signal that could not
 be
assigned to the four inequivalent sites in the structure. Hence, we
can only estimate an upper limit on the number of Cu(1)$_3$ sites
from our S/N and ultimately a lower limit on the chain length in
our sample.

\begin{figure}[tbh]
\resizebox{1.\linewidth}{!}{\includegraphics{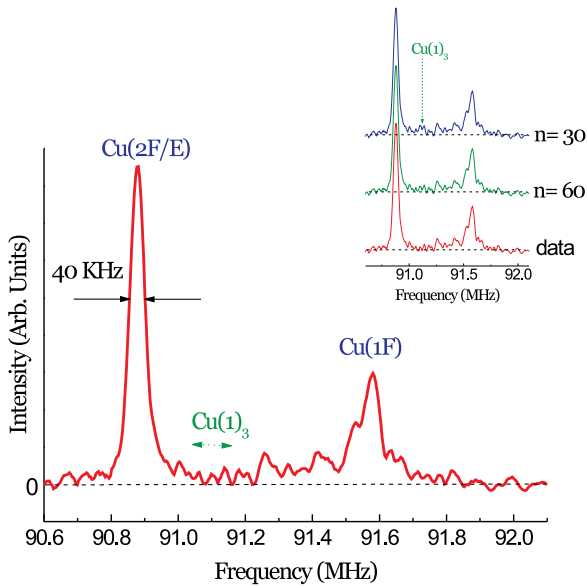}} 
\vskip -0.0cm \caption[]{ $^{63}$Cu NMR spectrum at 250 K and {\bf
H}$_0(=7.96 T)||${\bf c}.
The arrow shows the estimated range of the Cu(1)$_3$ resonance.
Inset illustrates presence of Cu(1)$_3$ line synthetically added
to the data for two chain lengths. This line is barely visible
for n=30 and undetectable for n=60.} \label{speccax}
\end{figure}

To put a quantitative limit on the concentration of Cu(1)$_3$
sites, we consider two sets of spectra. The first one was taken at
250 K where Cu(1F) is considerably narrower than the one at 70 K.
The Friedel--like oscillations are thermally smeared out at this high
temperature. In Fig. \ref{speccax} we have shown this spectra for
{\bf H}$_0||${\bf c}. We have used the established NMR parameters
of the Cu(1)$_3$ sites\cite{Lutgemeier92,Pieper} 
($\nu_{NQR}$ = 23.77 MHz, $\eta$ = 0.3 and $V_{zz}||${\bf b}) to calculate its
possible position. The horizontal arrow in Fig.~\ref{speccax}
covers the range consistent with parameter uncertainties. 
From the relative intensities of
Cu(1F), $I_4$, and Cu(1)$_3$, $I_3$,  one is able in principle to
find the mean number of oxygen atoms in the full chains (chain
length) according to the relation\cite{Lutgemeier92}:
n=1+2$I_4/I_3$. Since we have not observed any additional peaks
due to Cu(1)$_3$, we can only put a limit on the maximum intensity
of $I_3$ from our S$/$N. To obtain this limit, we have assumed the
Cu(1)$_3$ line has the same NQR width as observed by Yasuoka {\it
et al.} \cite{Yasuoka89} leading to an NMR linewidth of 66 kHz. If
we assume we can detect such a line with S$/$N of two then we
obtain a lower limit of 32 lattice constants for the chain length.
 The inset to Fig. \ref{speccax}
illustrates the Cu(1)$_3$ signal expected for n=30 and n=60.

The second set of spectra was taken at 60 K with {\bf H}$_0||${\bf
a}. In this case we use the fact that the relaxation rate of the
Cu(1)$_3$ site is relatively slow\cite{Vega} to extract only the
slowly relaxing components. As illustrated in Fig. \ref{spec1}, two
spectra taken at a repetition time of 25 msec. and 250 msec. have
been subtracted to reveal the Cu(1E) line as expected and no other
line. As above we put an upper limit on the chain length given our
S$/$N.  The calculated range for
Cu(1)$_3$ is displayed in Fig. \ref{spec1}. Our estimate of the
chain length from this analysis is 50 lattice constants.

\begin{figure}[tbh]
\resizebox{.8\linewidth}{!}{\includegraphics{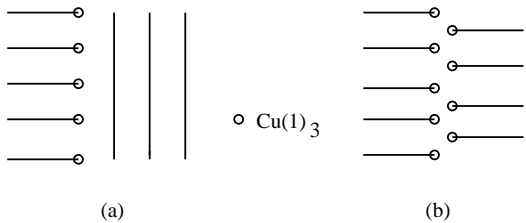}} 
\vskip -0.0cm \caption[]{ Domain interfaces calculated by Monte
Carlo calculations: (a) twin boundary and (b) parallel chains out
of phase. Line segments represent Cu--O--Cu chains. Chains are
terminated with Cu(1)$_3$ sites denoted with circles.}
\label{domainb}
\end{figure}

These estimates are consistent with the XRD result of 110 lattice
constants for the domain size along the chain direction. Using the
ASYNNNI model we have performed Monte Carlo calculations to determine
what the domain interface looks like for the two cases of (a) the
case of a twin boundary and (b) parallel chains that are out of
phase, {\it i.e.} FEFE adjoining EFEF (see Fig. \ref{domainb}).
In both cases a linear boundary is preferred as the ground state
with Cu(1)$_3$ atoms at the domain boundary. Thus we expect to
have three--fold coordinated copper sites at least along the domain
boundaries normal to {\bf b} and more along the twin boundaries to
the extent that they exist in our crystals. Hence the relative
concentration of Cu(1)$_3$ sites should be $I_3/I_4 \simeq$ 2\%
from the XRD results whereas our detection limit for $I_3/I_4
\simeq$ 4\%. From the two values for n obtained above, the consistent
estimate  
for the lower limit of the chain lengths is 50 lattice constants.

In NMR studies on HTSCs, large linewidths have been an obstacle
to investigating intrinsic properties. The problem arises from the
fact it is not readily obvious whether these large linewidths
arise from imperfections of the structure such as oxygen disorder
or are an intrinsic property of these materials. However, recent
improvements in sample preparation techniques have resulted in the
observation of narrower lines\cite{erb} emphasizing the importance
of sample quality in determining the properties of HTSC materials.
In general, the observed linewidths can be attributed to the
distributions in either the EFG parameters (arising from the
structure) and/or the Knight shift. Our analysis of the linewidths
of the central transitions vs. applied magnetic field as well
as a comparison between the linewidths of central and satellite
transitions indicates that the line broadening for Cu(2F/E) and
Cu(1E) sites is mainly caused by a distribution of EFG
parameters while for Cu(1F) it is mainly due to a distribution
of the Knight shift. The linewidths of all resonance lines
decreases with increasing temperature. The planar Cu sites for
{\bf H}$_0||${\bf c} have a linewidth of 40 kHz at 250 K. To our
knowledge these NMR linewidths are the smallest observed for
YBCO6.5 samples\cite{erb}, another indication of high sample
quality.

Oxygen disorder is a feature of non--stoichiometric samples. Erb
{\it et al.}\cite{erb} have compared the NMR spectra of YBCO7 and
YBCO6.9. The former has very narrow NMR lines. For example the
linewidth of  Cu(1)$_4$ is 22 kHz with {\bf H}$_0$ = 9 T $||${\bf c}
at room temperature. When the oxygen concentration is reduced by
0.1 the linewidth increases to 183 kHz, as a result of oxygen disorder.
Also noted was the presence of Cu(1)$_2$ with a linewidth of 13
kHz. They argue that this narrow linewidth is a result of oxygen
clustering which provides a more uniform environment for the
Cu(1)$_2$ sites than would be the case for randomly distributed
oxygen defects. They also argue that the absence of a Cu(1)$_3$
signal is consistent with oxygen clustering as this three--fold
coordinated site forms the boundary between full and empty chains.
As our linewidths are similar to that of YBCO7 we can infer that
there are few oxygen defects in our Ortho--II YBCO6.5 crystals. We
find that our Cu(1E) linewidth for fields along {\bf a} and {\bf
b} is 30 kHz at 200 K consistent with few oxygen defects. Our
narrow Cu(1F) line at these high temperatures is also consistent with
few oxygen defects. This is relevant to the Cu(1)$_3$ concentration.
Although we expect no substantial number of
oxygen defects in stoichiometric YBCO6.5 the absence of a
Cu(1)$_3$ line indicates that our crystals are indeed
substantially stoichiometric.

\begin{figure}[tbh]
\resizebox{1.\linewidth}{!}{\includegraphics{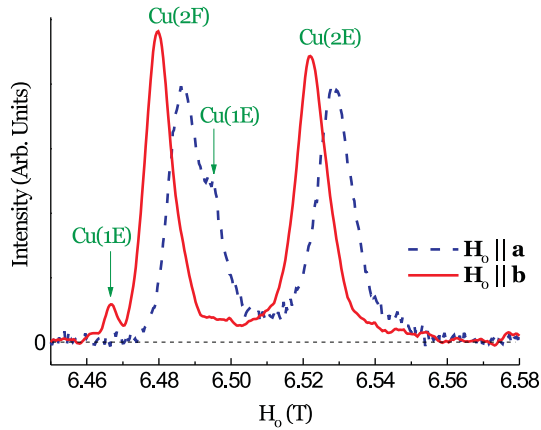}} 
\vskip -0.5cm \caption[]{ $^{63}$Cu NMR spectrum at 60 K and 75.75
MHz for {\bf H}$_0||${\bf a} and {\bf b}. The spectra are
normalized to the integrated intensity of Cu(2E) line. Note that
the feature on the right side of Cu(2F) for {\bf H}$_0||${\bf a}
is due to the imperfect subtraction of Cu(1E) signal. This of
course does not affect the Cu(2E) lineshape.} \label{twinning}
\end{figure}

From the NMR spectrum obtained for {\bf H}$_0||${\bf a} and {\bf
b}, we are also able to investigate the degree of twinning in our
crystal. Fig.~\ref{twinning} shows the $^{63}$Cu NMR spectrum for
Cu(2F) and Cu(2E) at 60 K, 75.75 MHz. A remnant of Cu(1E) is
observed on the high side of Cu(2F) as our pulse sequence to
saturate the Cu(1E) signal is not 100\% efficient. It is evident
from the lack of asymmetry of the two lines that there is no
obvious indication of the presence of twinning in our crystal.
Quantitatively we find by adding an admixture of the twinned
spectrum that to within our S$/$N, there could be a maximum of ten
percent twinning present in our sample. This is consistent with
our XRD measurements discussed above.

\subsection{Charge Transfer and Oxygen Ordering}

Oxygen ordering in the chain layer may have an effect
on charge transfer to the planes. We now investigate this 
possibility through the planar copper EFG values.
Considering the intinerant holes in the CuO$_2$ planes, $\nu_Q$ for the
two sites Cu(2E) and Cu(2F) in these planes might be expected to
be the same. However, we have observed a splitting of the Cu(2)
resonance. The coupling of the nuclei's
quadrupole moment with its environment determines $\nu_Q$ which is
directly proportional to the electric field gradient at the
nucleus. The local EFGs reflect the structural and
crystallographic aspects of the compound and provide information
about ionicity and bonding. The EFG at the nuclear site is
determined by two contributions: one arising from the charges on
the neighboring ions (lattice contribution) $V_{zz}^{lat}$ and
the second arising from the incomplete electronic shells (valence
contribution) $V_{zz}^{val}$ of the ion under consideration. The
total EFG may be written as\cite{sternheimer}

\begin{equation}\label{e:eq1}
V_{zz}= (1-\gamma_\infty)V_{zz}^{lat}+(1-R)V_{zz}^{val}
\end{equation}
where $\gamma_\infty$ and $R$ are the anitishielding factors which
account for the antishielding effect of the core electrons of the
central ion on the lattice and valence contributions,
respectively. From this equation, if one calculates the lattice
contribution, one can obtain the valence contribution from the
data. To what extent does the Cu(2) splitting stem from chain
ordering in the Ortho--II structure (lattice contribution) or from
differing charge transfer (valence contribution)? To address this
question, we have used the point charge model (PCM) to calculate the
lattice contributions to the observed EFG values for the four Cu
sites in Ortho--II structure. 
Although the PCM is simplistic
we are interested only in relative differences and not
the absolute values of EFG parameters.

Applying the formal valence description to Ortho--II YBCO results
in Y$^{+3}$Ba$_2^{+2}$Cu$_3^{+2}$O$_{6.5}^{-2}$ ionic
configuration. However this neglects the fact that Cu(1E) is in the
3d$^{10}$ state with charge +1. The resulting extra hole is
taken to be distributed over the CuO$_2$ planes.
We have taken the planar Cu sites to have a valence of +2.083,
the planar O sites a valence of -1.917, the Cu sites in the full
chain a valence of +2, the Cu in the empty chain a valence of +1,
the rest of the oxygens a valence of -2 and Ba and Y in +2 and +3,
respectively. We have used the atomic positions
obtained\cite{straube98} by single crystal X-ray diffraction
measurements at 80 K. It should be noticed that due to the
alternating nature of the full and empty chains, some atomic
displacements are observed\cite{straube98} from the mean position
of atoms, most notably Ba. This is attributed to the lattice
tendency to minimize its electrostatic energy\cite{zeiske}. We
have summed to convergence over the contributions within a crystal
containing $\pm$50, $\pm$100 and $\pm$30 unit cells along the $a$,
$b$, and $c$--axes, respectively. The resulting values for the
principal components of EFG for all sites are given in Table
\ref{t:efgcalc}. The (1-R)V$_{zz}^{val}$ values are obtained using
eq.~\ref{e:eq1}, the experimental values of $\nu_Q$ and
$\gamma_\infty$\cite{shimizu93} for
Cu$^{+2}$=-17.4 and Cu$^{+1}$=-5.2 in a crystalline environment.

\begin{table}
\caption{Calculated lattice EFG values, V$^{lat}$ in units of MHz, based on
the point charge model.}
\begin{ruledtabular}
\begin{tabular}{ccccccc}\label{t:efgcalc}
site   & V$_{aa}^{lat}$ & V$_{bb}^{lat}$ & V$_{cc}^{lat}$ &
$\eta^{lat}$& EFG-PA & (1-R)V$_{zz}^{val}$ \\\hline
Cu(2F) & -1.140 & -1.094   & 2.232   & 0.020     &c & -71.62  \\
Cu(2E) & -1.204 & -1.078   & 2.282   & 0.055     &c & -69.43  \\
Cu(1F) & 4.182  & -1.791   & -2.391  & 0.143     &a & -96.55  \\
Cu(1E) & 2.338  & 2.476    & -4.814  & 0.029     &c & -2.25    \\

\end{tabular}
\end{ruledtabular}
\end{table}

The calculated values produce the correct EFG principal axes.
Furthermore, small values of the asymmetry parameters for
Cu(2F$/$E) indicate a small deviation of these sites from
tetragonal symmetry in accordance with our experimental results.  The
directions of the principal axes and the values of $\eta$ are not
expected to be affected so much by the electronic structure of the
subject ion since this effect is isotropic to first order.
This explains the success of the point charge model
in predicting the principal axes and $\eta$ values.
A larger value of $\eta^{lat}$ is obtained for the Cu(1F) site.
This value is not as large as the observed one since it does not
contain the asymmetric ligand contribution.

Since the Cu(2F$/$E) and Cu(1F) sites have +2 valence, one would
expect\cite{EPRabragam} a valence contribution, (1-R)V$_{zz}^{val}
\approx -70$ MHz, to the EFG from a hole in the 3d$^9$ state and a
zero contribution for Cu(1E) sites in the 3d$^{10}$ state. Our
values obtained for this contribution to the EFG for Cu(1E) site
is small and supports this charge configuration for this site. We
find a difference in valence contribution of 2.19 MHz between the
Cu(2F) and Cu(2E) sites. Comparing this value to the
experimentally observed 3.15 MHz, we conclude that the valence
(local moment) contribution to the EFG difference at these sites
is dominant. This can be understood as resulting from a different
charge transfer for these two sites; the full chains act as the
charge reservoir whereas the empty chains do not. 
Charge transfered from the full chains resides at both the
Cu(2F) and Cu(2E) sites but with a larger
amount at Cu(2F), the site closer to the full chain.
The inadequacy
of the point charge model is apparent for the Cu(1F) site not only
in its prediction of the asymmetry parameter but also in the
valence contribution\cite{shimizuCu1F}. In this case asymmetric
ligand effects are mostly responsible. 

To investigate the effects of the domain walls on the EFG values,
we have introduced different types of domain walls and then
calculated the EFG parameters for different sites. This
calculation shows that the domain boundaries affect the Cu(1F$/$E)
sites the most and the Cu(2F$/$E) sites the least. Our results
also show that the bulk values of the lattice contribution to
$\nu_Q$ are essentially recovered just one lattice constant away
from the domain walls.

\section{Conclusions}\label{conclusion}

Line assignments for YBCO6.5 have been made. The lack of an observable NMR
signal from the chain ends at the domain walls confirms the long
chain lengths suggested by X--ray diffraction. We place a lower
limit on the chain length of 50 lattice constants.
Consistent with
this is the narrow linewidths comparable to those of the best
YBCO7 crystals. This is a result of little oxygen disorder
in these crystals.

Thus these crystals are a much better realization
of the stoichiometric YBCO6.5 than previously available.
Knight shift and spin--lattice relaxation measurements are
in progress to investigate the pseudogap. In addition,
as mentioned above, the long chains appear to support
Friedel--like oscillations nucleated at the chain ends\cite{zyamaniChain}.

Another consequence of oxygen ordering in the chain layer is the
effect it has on the CuO$_2$ planes via charge transfer. This
is manifested in the Cu(2E/F) splitting. Our analysis concludes
that most of this splitting is due to differential charge
transfer from the chains.
	
Finally, previous reports of static moments in YBCO6.5
are inconsistent with our results.

\acknowledgments We would like to thank J.P. Castellane and B.
Gaulin for performing the single crystal X--ray diffraction
measurements on our crystals. This work was supported by the
NSERC of Canada.

\appendix
\section*{Appendix}

\renewcommand{\theequation}{A.\arabic{equation}}

A straightforward calculation\cite{volkoff} to second order
reveals the transitions frequencies
for spin I=$\frac{3}{2}$:

\begin{equation}\label{e:nu}
 \nu  = \nu^{(0)} + \nu^{(1)}+ \nu^{(2)}
\end{equation}

\begin{equation}\label{e:nu0}
\nu^{(0)} = \gamma_n H_0 (1+K)
\end{equation}

\begin{equation}\label{e:nu1}
\nu^{(1)} =\frac{1}{2} \nu_Q (m-\frac{1}{2}) \left[ 3cos^2\theta
-1 - \eta sin^2\theta cos2\phi \right]
\end{equation}

\begin{widetext} \begin{equation}\label{e:nu2}
\begin{split}
\nu^{(2)}=& \frac{\nu_Q^2}{\gamma_n H_0 (1+K)} \bigg\{~
\frac{sin^2\theta }{32} \Big[ \{102m(m-1)-\frac{57}{2} \} \cos^2
\theta (1+\frac{2}{3} \eta cos 2\phi)\\
& \hspace{2.2cm}-\{6m(m-1)-\frac{9}{2} \}
(1-\frac{2}{3} \eta
cos2\phi) \Big] \\
& \hspace{2.2cm}+\frac{\eta^2}{72}
\Big[24m(m-1)-6-\{30m(m-1)-\frac{21}{2} \} cos^2\theta\\
& \hspace{2.2cm}- \{\frac{51}{2}m(m-1)-\frac{51}{8} \} cos^2 2\phi
(cos^2\theta-1)^2  \Big]~ \bigg\}
\end{split}
\end{equation}
\end{widetext}
where $K$ is the Knight shift obtained by
\begin{equation}\label{e:K}
K=K_x~ sin^2\theta cos^2\phi  + K_y~sin^2\theta sin^2\phi  +
K_z~cos^2 \theta
\end{equation}
and $\theta$ and $\phi$ are the polar angles of the applied
magnetic field with respect to the principal axis frame which is
assumed to be the same for both the Knight shift tensor and EFG
tensor. $\gamma_n$ is the nuclear gyromagnetic ratio and $H_0$ is
the applied magnetic filed. $\eta$ is the asymmetry parameter,
defined as $\eta=\frac{V_{xx}-V_{yy}}{V_{zz}}$, where
$|V_{xx}|\leq |V_{yy}| \leq |V_{zz}|$ are the principal components
of the electric field gradient (EFG) tensor at the Cu site and
$\nu_Q= eQV_{zz}/h$, $e$ the elementary electric charge, and
$Q$ is the electric quadrupole moment of the nucleus under study.
The central and satellite transition resonance frequencies are
obtained for $m$=$\frac{1}{2}$ and $m$=$\frac{3}{2}$, $-\frac{1}{2}$,
respectively.



\begin{thebibliography}{99}
\bibitem{T2comment} $T_2$ measured for Cu(2F) and Cu(2E) are the same.
\bibitem{Vega} Vega \etal\ \cite{Kwok88} and T. Imai, private
correspondance. Vega \etal\ have the Cu(1)$_3$ (23 MHz NQR line)
relaxation rate about three times faster than for Cu(1)$_2$ at
room temperature.
\bibitem{shimizuCu1F} It should also be mentioned that the
empirical values\cite{shimizu93} of $\gamma_\infty$ and $R$ for Cu ions are
obtained for sites with tetragonal symmetry. Hence, these values
might not be correct for Cu(1F) site.
\end{thebibliography}
\end{document}